# Ring-shaped velocity distribution functions in energy-dispersed structures formed at the boundaries of a proton stream injected into a transverse magnetic field: test-kinetic results

Gabriel Voitcu[1] and Marius Echim[1,2]

[1]Institute for Space Sciences, Atomistilor 409, Magurele 077125, Romania

[2]Belgian Institute for Space Aeronomy, Avenue Circulaire 3, Brussels B-1180, Belgium

In this paper we discuss the formation of ring-shaped and gyro-phase restricted velocity distribution functions at the edges of a cloud of protons injected into non-uniform distributions of the electromagnetic field. The velocity distribution function is reconstructed using the forward test-kinetic method. We consider two profiles of the electric field: (i) a non-uniform E-field obtained by solving the Laplace equation consistent with the conservation of the electric drift and (ii) a constant and uniform E-field. In both cases the magnetic field is similar to the solutions obtained for tangential discontinuities. The initial velocity distribution function is Liouville mapped along numerically integrated trajectories. The numerical results show the formation of an energy-dispersed structure due to the energy-dependent displacement of protons towards the edges of the cloud by the gradient-B drift. Another direct effect of the gradient-B drift is the formation of ring-shaped velocity distribution functions within the velocity-dispersed structure. Higher energy particles populate the edges of the proton beam while smaller energies are located in the core. Non-gyrotropic velocity distribution functions form on the front-side and trailing edge of the cloud; this effect is due to remote sensing of energetic particles with guiding centers inside the beam. The kinetic features revealed by the test-kinetic solutions have features similar to in-situ velocity distribution functions observed by Cluster satellites in the magnetotail, close to the neutral sheet.

I. INTRODUCTION

Experimental observations in various regions of the terrestrial magnetosphere and the magnetosheath have identified several classes of non-Maxwellian distributions as, for instance, ring shaped ion distributions[1], non-gyrotropic ion distributions[2–5] or unidirectional ion beams[6]. Their investigation is a topic of active research as the physical mechanisms responsible for the formation of such distribution functions are not fully understood. Frank *et al.*[2] suggest that non-Maxwellian distribution functions are the result of non-adiabatic acceleration of ions inside a current sheet. Meziane *et al.*[4] and Lee *et al.*[7] show that non-gyrotropic velocity distribution functions (VDFs) may form as a consequence of the remote sensing of a shock or a thin current sheet. Lee *et al.*[7] argue that in this case the computation of the moments of the distribution function is difficult and may provide unrealistically large bulk velocities. From a theoretical point of view, kinetic models of tangential discontinuities are constructed based on anisotropic velocity distribution functions that asymptotically tend towards isotropic and/or drifting Maxwellian functions.[8–11] As shown by Roth et al[10]., the Harris sheet, one of the models often used to describe current sheets, is a particular case of the generalized Vlasov equilibriums.

Energy-dispersed ion structures have been observed experimentally inside the terrestrial magnetosphere. For example, Bosqued *et al.*[12] and Zelenyi *et al.*[13] have used data from AUREOL 3 satellite to study energy-latitude dispersion relations for ion beams observed in the ionospheric high-latitude mapping of the plasma sheet boundary layer. More recently, Keiling *et al.*[14,15] identified multiple energy-dispersed ion structures in the plasma sheet boundary layer using multipoint measurements from the Cluster spacecraft. Various physical mechanisms have been proposed to explain this type of velocity-dispersed structures. Time-of-flight effects imprinted on the distribution functions by the electric drift of ions injected from the same source and moving with a reconnected frozen-in magnetic field line has been advocated by several authors.[13,16] Other mechanisms explain the observed velocity dispersion as a purely time-dependent process resulting from the satellite encounter with successive ion beams originating from the same source.[16] Other authors considered that the injection source itself has an energy-dependent structure.[14]

The test-kinetic simulation method provides a first approximation of the kinetic structure of a plasma system using prescribed electric and magnetic fields obtained from theoretical models, MHD simulations or experimental data.[17] It is a useful simulation tool especially in complex situations where the use of self-consistent kinetic methods is not practical. In the past, the test-kinetic method has been applied to study various problems of space plasma physics such as: mapping velocity distribution functions from the magnetosphere into the magnetosheath[18], studying plasma dynamics in the magnetotail and in the plasmasheet[19,20], investigating non-adiabatic effects introduced by sharp spatial variations of the electromagnetic field[21,22] or obtaining first order kinetic effects in collisionless perpendicular shocks in the vicinity of the Earth's bow shock[23,24].

In this paper we use the test-kinetic method to investigate the physical mechanisms responsible for the formation of energy-dispersed structures at the edges of a bounded proton cloud. We also discuss the formation of ring-shaped and non-gyrotropic velocity distribution functions within the energy-dispersed structure. We integrate numerically the trajectories of test-particles and use Liouville's theorem to map along them the initial velocity distribution function specified at the injection point. The VDF of protons is computed in various regions of the bounded cloud to investigate the effects due to nonhomogeneities in the electric and magnetic fields. The test-kinetic description is completely equivalent to a solution of the Boltzmann equation in the absence of collisions.[25] The novelty of our approach is to consider a bounded cloud/beam of protons and to investigate its kinetic structure in the regions close to its boundaries. We use two different configurations of the electric field and obtain similar results for the kinetic structure of the cloud boundary. Thus, the mechanism contributing to the formation of ring-shaped velocity distribution function and energy-dispersed structures is independent of the particular profile of the electromagnetic field. Indeed, we show that the gradient-B drift is responsible for the formation of the aforementioned kinetic features of the particle cloud.

The paper is organized as follows. In the second section we outline the main features of the test-kinetic method and describe how is it applies in our simulations. In the third section we illustrate the numerical solutions obtained for a cloud/beam of protons injected into a non-uniform electromagnetic field; we discuss the physical mechanisms that can explain the simulation results and can contribute to the formation of ring-shaped and energy-dispersed structures at the edges of the proton cloud/beam; we also compare the simulation results with experimental data from Cluster in the terrestrial magnetotail.[7] The last section includes our summary and conclusions.

## II. TEST-KINETIC MODELING

In a steady state situation the characteristics of the Vlasov equation can be obtained by solving the Newton-Lorentz equation of motion,

$$\frac{d\bm{v}}{dt} = \frac{q}{m}(\bm{E} + \bm{v} \times \bm{B}) \quad (1)$$

for an ensemble of charged particles injected into a stationary electromagnetic field distribution.[26] In Eq. (1) the electric field, $\bm{E}$, and the magnetic field, $\bm{B}$, are prescribed. One can compute any number of Vlasov characteristics and then "propagate" the VDF along them by applying the Liouville theorem. In this study we integrate numerically the equation of motion (1) for a large number of test-protons with initial velocities distributed according to a drifting Maxwellian.

In the test-kinetic approach the magnetic and electric fields introduced in Eq. (1) are prescribed a priori, particle trajectories are therefore not calculated fully self-consistently. In our computations the magnetic field is stationary and varies with the *x*-coordinate in a transition region centered at *x*=0. The B-field is confined everywhere in the *yOz* plane and it varies between two asymptotic fields, $\boldsymbol{B}_1$ on the left side of the discontinuity (*x*<<0) and $\boldsymbol{B}_2$ on the right (*x*>>0):

$$\boldsymbol{B}(x) = \frac{\boldsymbol{B}_1}{2} erfc\left(\frac{x}{L}\right) + \frac{\boldsymbol{B}_2}{2}\left[2 - erfc\left(\frac{x}{L}\right)\right] \quad (2)$$

where *L* represents the characteristic scale length of the transition region; *erfc* is the complementary error function. Note that similar profiles of the magnetic field are obtained self-consistently by kinetic models of one-dimensional tangential discontinuities.[9,10,27] The magnetic profile will be hereinafter also termed "discontinuity" although there is no discontinuous variation in the field.

In the simulations discussed in this paper we consider that the magnetic field is parallel to the *z*-axis. We superimpose a non-uniform two-dimensional electric field in the *xOy* plane (the *z* component vanishes identically). The E-field is everywhere normal to the magnetic induction $\boldsymbol{B}$ and it is obtained by solving the Laplace equation for the electric potential $\Phi$:

$$\nabla^2 \Phi = 0 \quad (3)$$

The solution domain is assumed to be a two-dimensional rectangular grid defined by the boundaries $[x_L, x_R]\times[y_B, y_T]$. Laplace's equation is complemented with Neumann boundary conditions:

$$\boldsymbol{n} \cdot \nabla \Phi = g(x) \quad (4)$$

where $\boldsymbol{n}$ is the unit normal vector of the boundary. The function *g* is specified by:

$$g(x) = \begin{cases} 0, & x = x_L \\ 0, & x = x_R \\ V_{0x} B_z(x), & y = y_B \\ -V_{0x} B_z(x), & y = y_T \end{cases} \quad (5)$$

where $V_{0x}$ is the plasma bulk velocity in the *Ox* direction on the left side of the discontinuity. The boundary conditions (5) have been defined such that the electric field in $y=y_B$ and $y=y_T$ sustaines a quasi-uniform electric drift in the direction normal to the surface of TD: $U_{Ex} = E_y(x)/B_z(x) = V_{0x}$. The boundary conditions at $x=x_L$ and $x=x_R$ correspond to a vanishing $E_x$ component at the two sides. The electric field obtained from Eq. (3) and Eq. (4) is a two-dimensional generalization of the 1D electric field used in previous test-particle simulations.[28] The solution of the Laplace equation (3) with boundary condition (4) is obtained numerically using the finite element method and it may be viewed as an electric field simulating the one sustained by space-charge layers forming at the boundaries of a moving non-diamagnetic plasma element in the presence of a magnetic field.[28–34]

The initial velocity distribution function of the protons injected on the left side of the TD (x<<0) is described by a drifting Maxwellian with the average velocity $V_0$ parallel to the positive *x*-axis and perpendicular to the discontinuity surface centered at *x*=0 :

$$f(v_x, v_y, v_z) = N_0 \left( \frac{m}{2\pi k_B T_0} \right)^{3/2} e^{-\frac{m\left[(v_x - V_0)^2 + v_y^2 + v_z^2\right]}{2k_B T_0}} \quad (6)$$

where $N_0$ and $T_0$ are the initial density and temperature of protons. The initial velocities of the test-particles are distributed according to Eq. (6). Particles are injected from $N_x \times N_y$ sources forming an uniform grid in the *xOy* plane separated by the distance $dx_0$ along the *x*-axis and $dy_0$ along the *y*-axis; *n* particles are injected from each source. A satisfactory sampling of the distribution function (6) is obtained for a total number of particles of the order of $10^6$.

The $6 \times n \times N_x \times N_y$ equations of motion are solved numerically in the time range [0, $\Delta t$], thus providing $3 \times n \times N_x \times N_y$ components of the test-particles velocities at moment $\Delta t$. These final velocities define a non-uniform 3D mesh in the velocity space. The direct test-kinetic approach, or the Liouville mapping, consists in assigning to each node of the mesh defined by the final velocities, $(v_x^i, v_y^i, v_z^i)$, the value of the VDF computed from Eq. (6) for the initial velocity components $(v_{x0}^i, v_{y0}^i, v_{z0}^i)$; this is done accordingly for each particle *i* of the ensemble. This procedure is applied for those spatial bins of the physical space populated by a large number of particles where the velocity distribution function is sampled rigorously. The numerical method used to integrate the trajectories of the test-particles is based on a Cash-Karp Runge-Kutta algorithm with adaptive step-size and has been parallelized to increase the computation speed.

## III. EXPERIMENTAL AND NUMERICAL EVIDENCE FOR RING-SHAPED AND NON-GYROTROPIC VELOCITY DISTRIBUTION FUNCTIONS

### A. Ring-shaped and non-gyrotropic VDFs observed by Cluster in the terrestrial magnetosphere

Lee *at al.*[7] and Wilber *et al.*[5] have thoroughly analyzed data from the four Cluster spacecrafts[35] on 1st of October 2001 and discussed the kinetic properties of the plasma in the vicinity of the neutral sheet. It is shown that the current sheet became very thin, of the order of the Larmor radius of a 5 keV proton gyrating into a magnetic field of 10 nT. The thinning of the current sheet coincided with observations of highly anisotropic ion velocity distribution functions. $B_{y,\text{GSM}}$ is turning from negative to positive values, as discussed by Lee *et al.*[7] Examples of ion distribution functions corresponding to this time interval are illustrated in Fig. 1 showing projections in the space of perpendicular velocities.

Before the thinning of the current sheet, the proton velocity distribution functions observed by Cluster spacecraft (see Fig. 1) are nearly isotropic with a small central cavity. Nevertheless, later on, the velocity distribution function, although remains still gyrotropic, shows that a significantly larger central region of the velocity space is voided of particles. The distribution function becomes highly non-gyrotropic toward the end of the analyzed time interval, with the ions being restricted to a limited sector in the space of perpendicular velocities. Similar features of velocity distribution functions have been reported by observations of ion shell or ring-shaped velocity distribution functions in the plasma sheet[36] or plasma sheet boundary layer – PSBL[37], triggering instabilities, waves and auroral emissions.[38–40]

Based on test-particles simulations without any electric field, Lee *et al.*[7] have been able to explain the observations of the crescent-shaped non-gyrotropic velocity distribution functions by the remote sensing of particles with Larmor radius larger than the thickness of the neutral sheet. The velocities of the test-particles have been initialized by Lee *et al.*[7] according to a Maxwellian distribution function with a low-energy cut-off matching the observed velocity distribution functions. The VDF in various regions of the simulation domain has been computed by Lee *et al.*[7] by accumulating particles passing through the corresponding spatial cell in 25 ion Larmor periods and not by tracing VDFs along integrated orbits as in this study.

Inspired by this example of experimental and numerical results we performed test-kinetic simulations of the magnetic configuration described by Eq. (2), similar to the one imagined by Lee *et al.*[7] The test-kinetic method adopted here enables the direct computation of the velocity distribution function by Liouville mapping. We consider an antiparallel profile of the magnetic field such that ***B***, described by Eq. (1), is everywhere parallel to $z$-axis and changes sign at $x$=0 (see Fig. 2). Fig. 3 shows the electric field distribution computed from Laplace's equation (3) subject to boundary conditions (4) discussed in the previous section. This distribution may be viewed as describing a neutral sheet and a superimposed electric field with $E_y$ changing sign whenever the $B_z$ reverses sign. One important aspect of our electromagnetic configuration is that although $B_z$ changes sign at $x$=0, the gradient-B drift remains parallel to +$Oy$ since $\nabla B$ also changes sign at $x$=0. A schematic diagram showing the simulation geometry is given in Fig. 4.

**B. CASE I: antiparallel magnetic field and non-uniform electric field**

Our numerical experiment consists in injecting test-particles on the left side of the neutral sheet (x<0, see Fig. 2) and integrating their orbits until they intersect the magnetic transition region (or discontinuity). The initial velocity distribution function of the test-particles is given by the drifting Maxwellian (6) without any low-energy cut-off. All the trajectories are integrated numerically over a time interval $\Delta t$. The simulation domain is defined between [−40000, +40000] km along the *x*-axis and [−30000, +30000] km along the *y*-axis, the particles which reach regions outside these limits are removed from simulation. The input parameters assumed for the simulations are given in Table I.

In Fig. 5 we illustrate the positions of the protons in the *xOy* plane, perpendicular to the magnetic field, after 120 seconds of gyration and drift (roughly 55 gyration periods). The local number density is color coded; one density value is assigned to each bin of a 2D mesh of 60x60 spatial cells. After 120 seconds the protons are still in the region with a smooth variation of the magnetic and electric fields, on the left side of the discontinuity, and continue to drift towards the discontinuity. The spatial variation of the velocity distribution function in various regions of the cloud is shown in Fig. 6. As expected, in the central region of the proton cloud the VDF is a drifting Maxwellian close to the one initialized at *t*=0.

After a sufficiently long time interval the protons drift inside the transition region where they interact with the non-uniform field. Their positions in the *xOy* plane, perpendicular to the magnetic field, are shown in Fig. 7 for $\Delta t$ = 225 seconds (or roughly 100 proton gyroperiods). The overall shape of the proton cloud is deformed and shows significant asymmetries; the test-protons are scattered in the positive direction of the *y*-axis. While initially the spatial scale of the proton cloud in the *y*-direction was roughly 6000 km, after 225 seconds the cloud expands over 20000 km in the positive direction of *y*-axis. We assign this asymmetric expansion of the cloud to the positive gradient-B drift[41]:

$$V_{\nabla B} = \frac{m w_\perp^2}{2qB^3} \boldsymbol{B} \times \nabla B \quad (7)$$

where $w_\perp$ is the perpendicular velocity in the guiding center frame. Indeed, for the geometry chosen here the drift described by Eq. (7) is the most important one. This first order charge dependent gradient-B drift acts inside the transition region where the electromagnetic field is non-uniform. The velocity distribution function of protons in various spatial bins inside the cloud is computed for each spatial bin defined in the *xOy* plane and identified by the combination of letters (rows) and numbers (columns) in Fig. 7. The size of a spatial bin is defined such that it contains enough particles for a good sampling of the velocity space. The bins of the mesh shown in Fig. 7 have a spatial resolution of 280 km in *x*-direction and 2500 km in *y*-direction, adapted to the geometry of the cloud and the total number of simulated particles.

The spatial variation of the velocity distribution function for each bin of Fig. 7 is shown in Fig. 8. These results illustrate a key-feature of the proton velocity distribution function: the formation of a cavity in the central region of the perpendicular velocities space. An example is given by the ring-shaped velocity distribution functions obtained in the lateral spatial bins D3, E3, D4 and E4. The size of the central, low-energy cavity has the tendency to increase with the distance from the center of the ion cloud; the central void of particles is larger close to the upper boundary (i.e. for larger *y*-values) than in the center (compare, for instance, the velocity distribution functions of bins E3 and E4). The kinetic properties of the front and trailing edge of the proton cloud show additional interesting features. The velocity distribution function is highly non-gyrotropic on the front-side and trailing edge of the cloud, as illustrated by the VDFs obtained for instance in spatial bins A4, B4, G4 and H4, shown in Fig. 8. These aspects will be discussed in more detail below.

Further, Fig. 9, 11, 13 and 15 shows the positions of the protons in the *xOy* plane after 275, 300, 350 and 600 seconds of gyration and drift illustrating the subsequent stages of the interaction of the proton cloud/beam with the region of magnetic field gradient. The proton cloud continues to drift towards the right side of the discontinuity. The overall shape of the cloud is strongly deformed and shows significant asymmetries. The spatial variation of the velocity distribution functions corresponding to the cloud's position illustrated by Fig. 9, 11, 13 and 15 are shown in Fig. 10 for $\Delta t$=275 s, Fig. 12 for $\Delta t$=300 s, Fig. 14 for $\Delta t$=350 s and Fig. 16 for $\Delta t$=600 s. The results confirm that the ring-shaped distribution functions observed at the edges of the proton cloud/beam are preserved even when the cloud exit outside the region of magnetic transition.

Fig. 13 illustrates the formation of two proton populations: P1, a population which is captured in the interior of the discontinuity and P2 a population that penetrates inside the discontinuity and moves across it into the right side. The formation of these two populations is an effect due to the combination of the electric drift and gradient-B drift acting inside the simulation domain. A detailed analysis of the trapped and penetrating populations will be published in a subsequent paper; here we emphasize mainly the kinetic properties of the boundaries of the proton cloud/beam.

As shown by Fig. 8, 10, 12, 14 and 16 the lateral edges of the cloud (in the direction perpendicular both to *V* and *B*) are mainly populated by the most energetic particles of the cloud and the velocity distribution functions in these regions are ring-shaped. We conjecture that the physical mechanism that can explain the formation of the central cavity in the velocity space and the creation of a ring-shaped distribution in the space of perpendicular velocities is related to the expansion of the cloud in the +$Oy$ direction, thus it is an effect due to the gradient-B drift. Indeed, Eq. (7) shows that the particles with a larger perpendicular thermal energy are deflected by $V_{\nabla B}$ to larger distances along the $y$-direction. The velocities of the particles deflected by the gradient-B drift populate the ring shaped VDFs observed, e.g., in bins D3, D4, D5, E3, E4, E5, F4, F5 of Fig. 8. The protons with smaller thermal velocities are deflected less by $V_{\nabla B}$ and therefore their number decreases in the positive direction of the $y$-axis, away from the center of the cloud. The regions close to the edges are virtually not accessible to particles with small thermal energy. Thus, the velocity space corresponding to the smaller energies is void and a cavity is formed in the central part of the velocity distribution function (see bins D3, D4, D5, E3, E4, E5, F4, F5 of Fig. 8).

An additional key-feature associated with the same physical process is the formation of a velocity-dispersion structure at the edges of the cloud. Since $V_{\nabla B}$ displaces most efficiently the particles with higher thermal velocities, the outer edges of the proton cloud/beam will be populated by the particles with the highest energies. Thus the kinetic energy of the protons decreases from the outer layers towards the core of the cloud. This velocity dispersion is revealed for instance by the VDFs in bins F5, F4, F3, F2, F1 of Fig. 8. Our simulations results suggest an additional mechanism[13,14,16] to explain the formation of velocity-dispersed ion structures which is based on the gradient-B drift that can imprint energy-dispersed features in regions close to the boundaries of spatially confined plasmas.

The kinetic properties of the front-side and trailing edge of the proton cloud show interesting additional kinetic features. The proton velocity distribution function is non-gyrotropic in these regions and it has a crescent shape with increased phase density in certain regions of the perpendicular velocities space. This effect is seen, for instance, in bins A3, A4, H2-H5 of Fig. 8, A3, A4, H3, H4, I4 and I5 of Fig. 10, A2, A3, H3 and I3 of Fig. 12, A2, A3, D1, E2, F3, G3, H4, I4 of Fig. 14. These crescent-like shaped VDFs observed at the front and trailing edges of the cloud are an effect due to the remote sensing of the protons with guiding centers localized in the interior of the cloud and whose Larmor radius is large enough to enable the orbit to penetrate along $x$ in regions behind and ahead of the bulk of the cloud. For example, most of the particles localized in bins D1, E2, G3 and I4 from Fig. 14 have a negative gyration velocity $v_y$ due to the clockwise gyration of protons in the positive B-field. At the same time the particles in bins A3, B4, C5 shown in Fig. 14 have a positive gyration velocity $v_y$ due also to the clockwise gyration. Therefore, the gyro-phase restricted effect observed on the front-side and trailing edge region can be associated with the remote sensing of particles whose guiding centers pertain to the interior of the cloud. Similar conjectures have been made by Lee et al.[7] for in-situ data from the magnetospheric tail and by Meziane et al.[4] for high energy ions near the terrestrial foreshock.

**C. CASE II: parallel magnetic field and uniform electric field**

In order to confirm the conjectures made on the effects of the gradient-B drift on the formation of ring-shaped and non-gyrotropic velocity distribution functions we carried out a second set of test-kinetic simulations for a unidirectional, non-uniform and increasing magnetic field and a uniform distribution of the electric field. The magnetic field is everywhere parallel to the positive $z$-axis and its magnitude increases from 30 to 90 nT without changing of sign. The electric field is computed as a convection field determined by the orientation and the magnitude of the asymptotic magnetic field $\boldsymbol{B}_1$ and the bulk velocity $\boldsymbol{V}_0$ of plasma on the left side of the transition region:

$$\boldsymbol{E} = \boldsymbol{B}_1 \times \boldsymbol{V}_0 = \text{const.} \tag{8}$$

Thus, in case II the electric field is everywhere parallel to the positive $y$-axis. The input parameters assumed for this simulation are given in Table I. The position of the test-protons in the $xOy$ plane, at the end of the simulation, is shown in Fig. 17. The local number density is color coded using a spatial mesh of 80x80 bins. Fig. 17 shows that the protons are scattered in the direction of the gradient-B drift, i.e. the positive $y$-axis. Fig. 18 illustrates the spatial variation of the velocity distribution function of protons. The spatial resolution used to compute the VDF is equal to 150 km along $x$-direction and 3000 km along $y$-direction. The results show that ring-shaped VDFs similar to the ones obtained in case I are obtained in bins D3, D4, E3, E4, F3 and F4. The size of the central cavity formed in the perpendicular velocities space increases as we approach to the upper edge of the cloud. Crescent-like, non-gyrotropic distribution functions are obtained on the front and trailing edge of the cloud, similar to case I (see for instance the VDFs corresponding to the bins I1 to I5 and to the bins A1 to A5). The results obtained in case II, for a parallel magnetic field and a uniform electric field confirm the role of the gradient-B drift on the formation of non-Maxwellian distribution functions.

The simulations included in this paper are based on input parameters (velocity distribution functions, magnetic field) consistent with experimental data from the terrestrial magnetosphere and emphasize the role of the gradient-B drift at sharp transitions to contribute to the formation of non-Maxwellian distribution functions (ring-shaped and crescent-like) as those observed by Cluster in the plasma sheet. The velocity distribution functions obtained numerically show similarities with those observed experimentally by Cluster satellites in the terrestrial magnetotail. We compared the proton velocity distribution functions measured by Cluster CIS instruments and shown in Fig. 1, with the proton velocity distribution functions computed numerically using our test-kinetic approach, illustrated in Fig. 14 and 18. Both measured and computed VDFs have cavities in the central region of the perpendicular velocities space, like those illustrated by the left and middle panels of Fig. 1 and respectively by the VDFs corresponding to the spatial bins C2 and D3 from Fig. 14. In our simulations the central cavity results as a consequence of a physical process related to the spatial dispersion of particles due to the gradient-B drift and is not due to an ad-hoc low energy cut-off imposed onto the initial velocity distribution function of the test-particles. On the other hand, crescent-like, non-gyrotropic VDFs are revealed by satellite data (right panel from Fig. 1) and also by our simulated data (e.g. bins E2 and G3 from Fig. 14, or bins H4-H5 and I4-I5 from Fig. 18). This effect is a consequence of the remote sensing in regions outside the cloud of particles whose guiding centers are found inside the cloud.

## IV. SUMMARY AND CONCLUSIONS

In this paper we have imagined a configuration of the electromagnetic field that enables the investigation of the dynamics of a proton cloud/beam injected across sheared electric and magnetic fields. The parallel component of the electric field is everywhere equal to zero and the perpendicular component is computed by solving Laplace's equation on a two-dimensional rectangular grid (case I) or is constant and uniform (case II). We analysed the individual motion of particles and the Liouville mapping of an initial velocity distribution function of the protons. The test-particles move across regions with sharp variations of a sheared antiparallel (case I) or parallel (case II) magnetic field. The global dynamics of the cloud is asymmetric due to the gradient of B drift that is oriented in the $+Oy$ direction. The asymmetry is imprinted on the cloud morphology while it traverses the transition region. However, the cloud remains asymmetric at large distances from the magnetic transition region, in the region of uniform field.

The asymmetry of the cloud is retrieved in its kinetic structure. Indeed, the layer formed at the outer edge of the cloud, in the positive *y*-direction perpendicular to the bulk velocity and magnetic field, is populated by particles whose velocities varies with the distance from the cloud's core, forming an energy-dispersed structure. This kinetic feature is obtained in both simulations, case I and II, when the magnetic field profile is parallel and antiparallel. The test-kinetic simulations illustrate that the energy of the particles increases towards the fringe of the cloud. We have shown that this effect is due to the gradient-B drift that efficiently disperses protons in the $+Oy$ direction, proportionally to their kinetic energy.

The reconstruction of the velocity distribution function by the direct (or forward) test-kinetic approach shows the formation of a central cavity in the space of the perpendicular velocities. Such ring-shaped distribution functions are obtained in spatial bins localised close or within the outer energy-dispersed layer.

Since the particles with a smaller thermal energy are less deflected they populate mainly the core of the cloud at lower $y$-coordinates where the velocity distribution function is close to a Maxwellian. On the front-side and in the trailing edge of the cloud the velocity distribution function is non-gyrotropic and has a crescent like shape in some of the spatial bins. This anisotropy is an effect due to the remote sensing of particles whose guiding centers pertain to the inner cloud.

One of the advantages of the direct test-kinetic method, i.e. forward integration in time of the particle trajectories, is that it enables the investigation of the VDFs and also gives an idea of the general dynamics of the cloud while convecting in the non-uniform field configuration. A disadvantage of the direct method is that the velocity distribution function is reconstructed for spatial bins with rather coarse dimensions imposed by the total number of simulated particles.

The main conclusions of our study do not necessarily depend on the particular profile of the electric field nor on the existence of a mechanism able to inject clouds of ions in the neutral sheet. This is confirmed by the results obtained in case II where the electric field is uniform. Our simulations have been performed for a configuration that reproduces some typical parameters for the terrestrial plasma sheet investigated previously by Lee *et al.*[7] and Wilber *et al.*[5] In our simulations, however, we introduce an electric field and a finite sized cloud of test-particles. A possible origin of the electric field treated in our simulations may be the launching of Bursty Bulk Flows (BBFs) in the plasma sheet and considering their propagation across the neutral sheet[42], or the ballooning instability[43], or the propagation of bubbles in the geomagnetic tail related also to BBFs formation[44]. Testing of any of these mechanisms is beyond the scope of our paper. But in all the three scenarios mentioned above a localised perturbation of the dawn-dusk, cross-tail, electric field is observed and can be considered as a possible source for the prescribed electric field used in simulations.

Although we did not show the projections of the VDFs in the plane of velocities including the parallel direction, the phase space density is equally distributed in the parallel and anti-parallel direction of the magnetic field. The proton cloud expands along the magnetic field lines due to the parallel velocity component assigned initially. In a realistic magnetospheric configuration the particles with the larger positive parallel velocities moving along $B$, populating the upper half of the cloud, will be reflected by the mirror force at some ionospheric altitude and will travel back thus filling some regions of the negative parallel velocity space. However these effects have not been investigated yet by our simulations.

The formation of the energy-dispersed structure and of the ring-shaped velocity distribution function is a kinetic effect obtained at the edges of a localized plasma cloud. In a magnetotail geometry when the cloud/stream moves in the $z_{GSE}$ direction, the B-field being mainly along the $x_{GSE}$, the ring-shaped VDFs and the energy-dispersed structure will be observed at the lateral edges in the $y_{GSE}$ direction. Thus, the localisation of the non-Maxwellian velocity distribution functions and possibly their properties would depend on the local magnetic field and the global geometry of the cloud itself. In a future study we shall investigate in detail this relationship.

The test-kinetic simulations discussed in this paper suggest a physical mechanism that can explain the formation of an energy-dispersed structure at the edges of a proton beam interacting with a non-uniform electromagnetic field. We have also identified kinetic effects contributing to the formation of ring-shaped and non-gyrotropic velocity distribution functions. Although these results are obtained for two prescribed configurations of the electromagnetic field their relevance is more general. Our study emphasizes the role of the gradient-B drift on the edges of plasma structures; the gradient-B drift is acting in any region with a non-uniform magnetic field, like for instance, X-lines or X-planes. In more complex geometries and for time dependent situations additional first order drifts may be included. Nevertheless, the main results of our study should remain valid as all the first order drifts depend on the kinetic energy of the particle. This is a key-property emphasized by the kinetic effects described in this paper. A preliminary comparison with in-situ data from the terrestrial magnetosphere show a rather good correlation and more in-depth data comparisons will be done in the future.

**ACKNOWLEDGEMENTS**

This paper is supported by the Sectoral Operational Programme Human Resources Development (SOP HRD) financed from the European Social Fund and by the Romanian Government under the contract number SOP HRD/107/1.5/S/82514. Also, the authors acknowledge support from the Solar Terrestrial Center of Excelence (STCE) in Uccle, Brussels, Belgium, from the European Space Agency (ESA PECS Project 98049 - KEEV) and from the International Space Science Institute in Bern, Switzerland through the project "Plasma Entry and Transport in the Plasma Sheet" led by Simon Wing and Jay Johnson. The authors are also grateful to Richard Marchand for reading and commenting a draft of this paper. Marius Echim thanks Joseph Lemaire for many enlightening discussions about plasma dynamics in non-uniform electromagnetic fields.

Table I. Input parameters of the test-kinetic simulations: $N_0$, $kT_0$, $V_0$ are the density, thermal energy and average velocity of the drifting Maxwellian given by Eq. (6); $B_{1z}$, $B_{2z}$ are the asymptotic values of the magnetic field; $L$ is the length scale of the discontinuity; $R_L$ is the Larmor radius of the thermal protons in the frame of reference moving with $V_0$; $T_L$ is the proton cyclotron period; $N_x \times N_y$ is the number of injection sources; $n$ is the number of test-particles injected from each source; $x_0$, $y_0$ are the coordinates of the first source; $dx_0$, $dy_0$ are the separation distances between sources along $Ox$ and $Oy$.

|  | $N_0$ [m$^{-3}$] | $kT_0$ [keV] | $V_0$ [km/s] | $B_{1z}$ [nT] | $B_{2z}$ [nT] | $L$ [km] | $R_L$ [km] | $T_L$ [s] | $N_x, N_y$ | $n$ | $x_0, y_0$ [km] | $dx_0, dy_0$ [km] |
|---|---|---|---|---|---|---|---|---|---|---|---|---|
| **Case I** H$^+$ | $10^4$ | 3 | 200 | −30 | +30 | 6000 | 260 | 2.2 | 12, 12 | 20000 | −20000, −550 | 200, 100 |
| **Case II** H$^+$ | $10^4$ | 7 | 200 | +30 | +90 | 6000 | 400 | 2.2 | 12, 12 | 75000 | −20000, −1100 | 200, 200 |

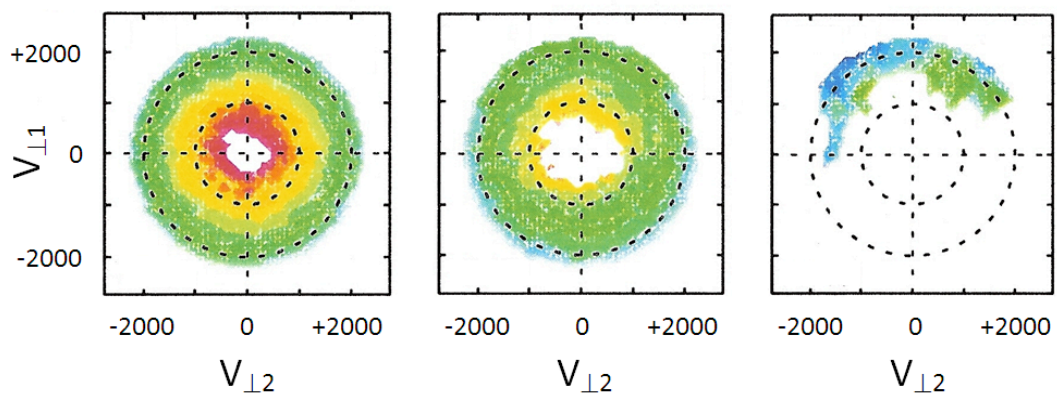

Fig. 1. Example of ring-shaped and non-gyrotropic (crescent-like) ion velocity distribution functions observed by Cluster (C1) CIS instruments[45] at 09:25:40 UT (left panel), 09:36:32 UT (middle panel) and 09:46:11 UT (right panel) on 1st of October 2001.[7] All panels illustrate sections in the velocity plane perpendicular to the local magnetic field; the units are km/s on both axes (from Lee *et al.*[7]).

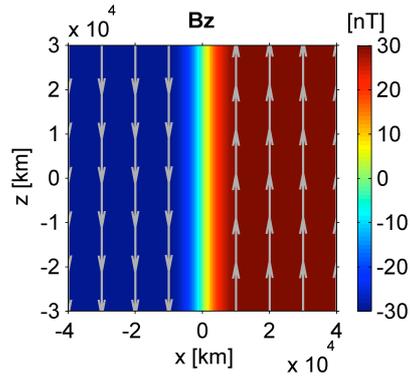

Fig. 2. Magnetic field distribution in the simulation domain for case I. The field is unidirectional and changes orientation at *x*=0.

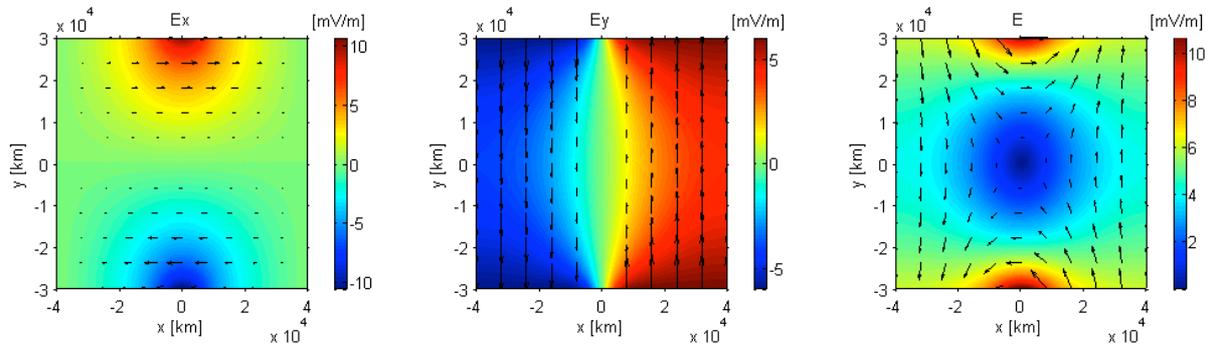

Fig. 3. Electric field distribution in the simulation domain for case I: (left panel) $E_x$ component, (middle panel) $E_y$ component, (right panel) magnitude of the E-field. The integration domain is defined by: [−40000,+40000] km × [−30000,+30000] km.

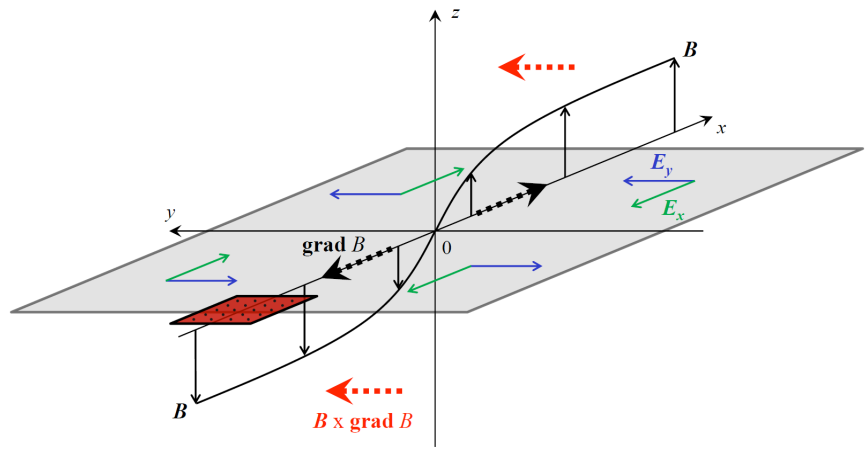

Fig. 4. Schematic diagram of the simulation geometry.

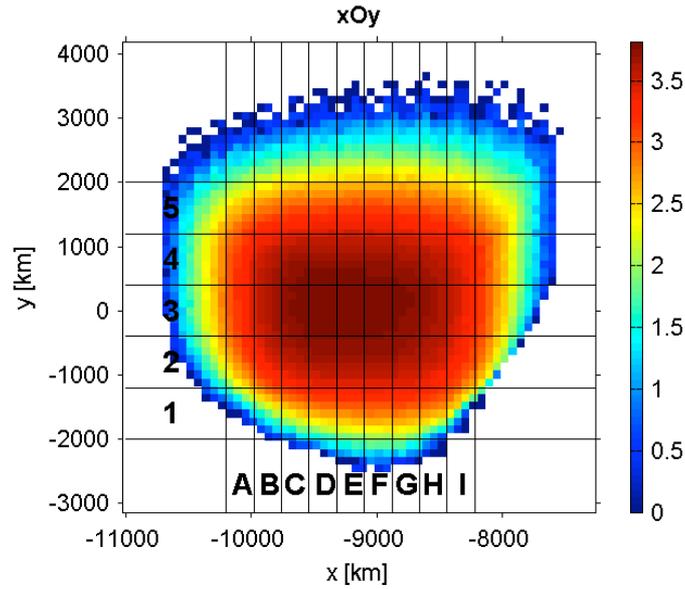

Fig. 5. Distribution of protons in the *xOy* plane after 120 seconds (~55$T_L$) from injection in the electromagnetic field illustrated in Figures 2 and 3. The local value of the number density is color coded. The cloud moved in a region of virtually uniform electric and magnetic field. The spatial mesh on which the VDF is reconstructed is also shown; each bin is identified by a combination of letters and numbers as shown on the left side and bottom side of the figure.

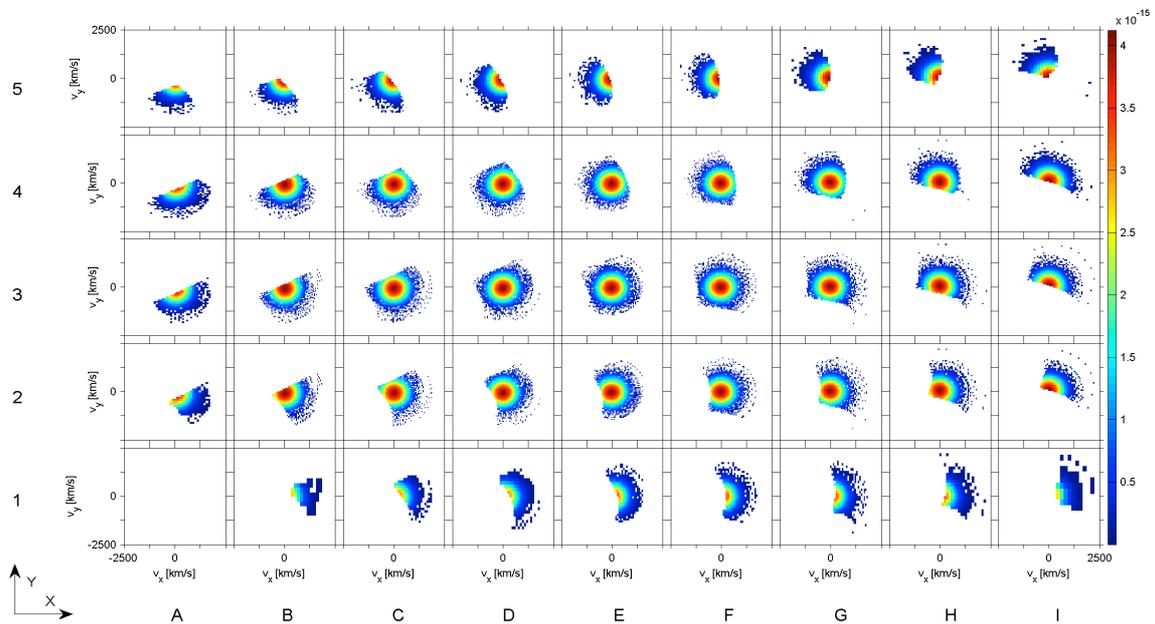

Fig. 6. Projection in the space of perpendicular velocities, for $v_z$=0, of the Liouville mapped velocity distribution functions of protons in the cloud at Δ*t*=120 seconds. The spatial bins are defined in Fig. 5. One notes the drifting Maxwellian VDF obtained in the central core of the cloud and non-gyrotropic VDFs at the edges of the cloud; the latter result from large Larmor radius particle whose gyrocenters are inside the cloud. Note the different regions of the velocity space populated in bins C1–I1 compared to C5–I5.

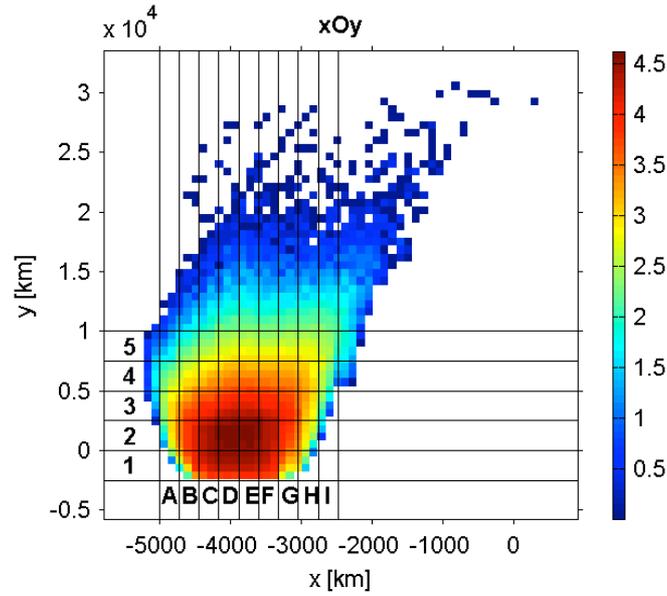

Fig. 7. Distribution of protons in the *xOy* plane after 225 seconds (~100$T_L$) from injection in the electromagnetic field illustrated in Figures 2 and 3. The local value of the number density is color coded. The cloud has spent some time in the region of non-uniform fields and its shape is elongated into the +*y*-direction under the action of the gradient-B drift. The spatial mesh on which the VDF is reconstructed is also shown.

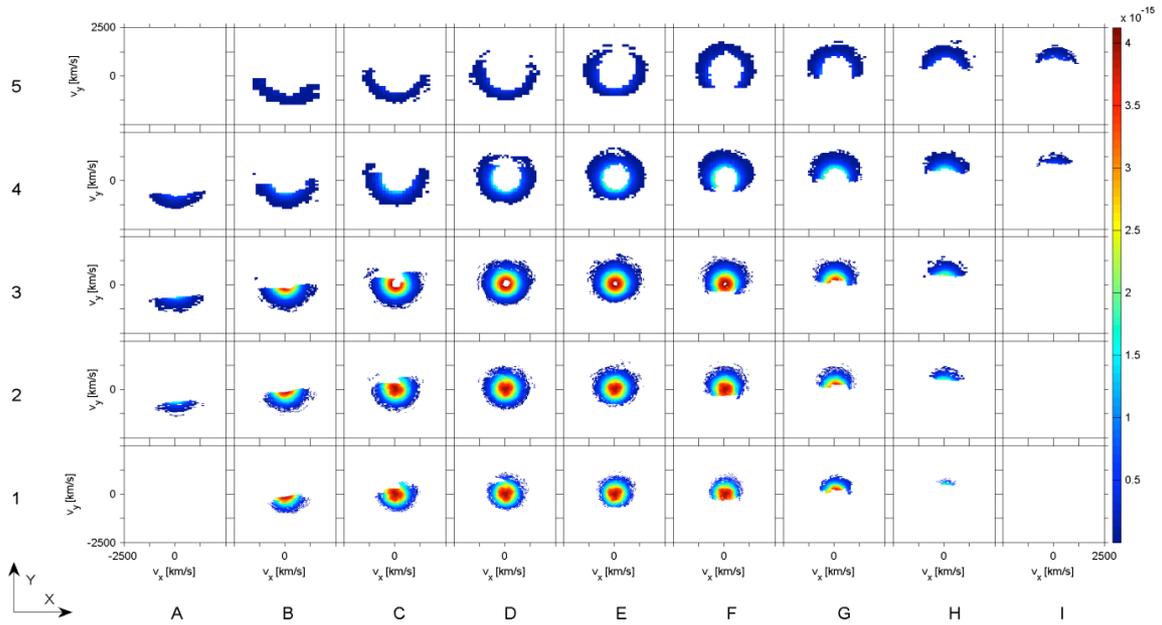

Fig. 8. Projection in the space of perpendicular velocities, for $v_z=0$, of the Liouville mapped velocity distribution functions of protons in the cloud at $\Delta t$=225 seconds. The spatial bins are defined in Fig. 7. A drifting Maxwellian VDF is obtained in the core of the cloud, in bins D1−F1, D2−F2. Ring-shaped VDFs are obtained close to the edges of the cloud, bins D4−F4, E5−F5. Crescent-like VDF are obtained close to the trailing and leading edges, A2−A4, B4−B5, C4−C5, G4−G5, H4−H5, I4−I5.

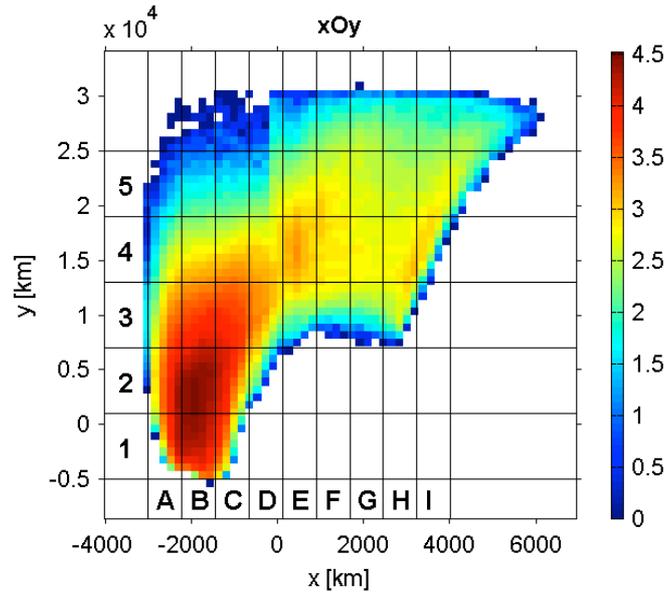

Fig. 9. Distribution of protons in the *xOy* plane after 275 seconds (~125$T_L$) from injection in the electromagnetic field illustrated in Figures 2 and 3. This snapshot illustrates the initial stage of the interaction between the proton cloud and the region of the most rapid variation of **B**; some parts of the cloud intersected the plane *x*=0 where **B**=0. The local value of the number density is color coded. The spatial mesh on which the VDF is reconstructed is also shown.

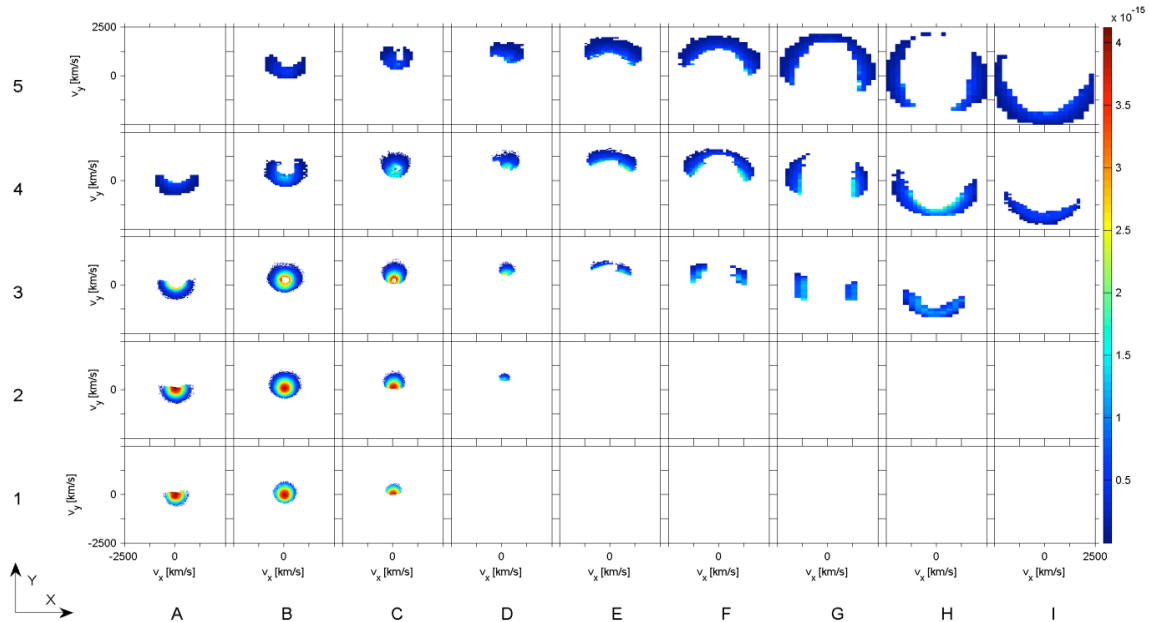

Fig. 10. Projection in the space of perpendicular velocities, for $v_z$=0, of the Liouville mapped velocity distribution functions of protons in the cloud at Δ*t*=275 seconds. The spatial bins are defined in Fig. 9. During this initial stage of the interaction of the cloud with the discontinuity one identifies the Maxwellian core of the cloud (bins B1–B2) and non-gyrotropic VDFs at the leading edge (e.g. bins D5–I5).

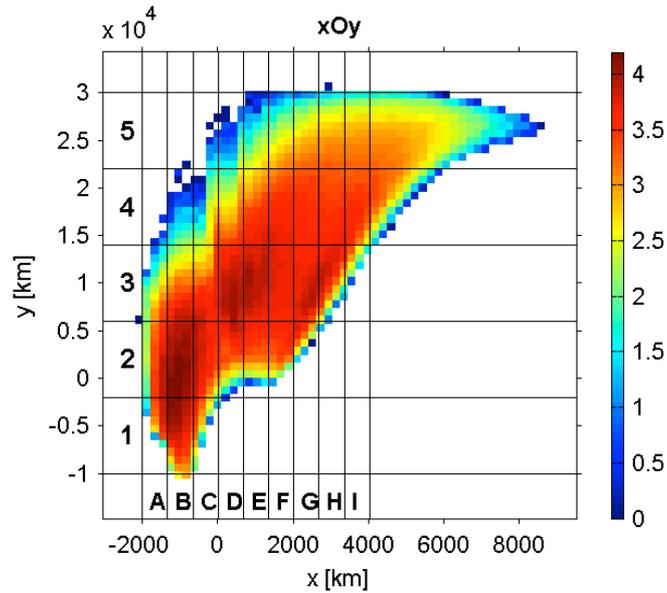

Fig. 11. Distribution of protons in the *xOy* plane after 300 seconds (~$135T_L$) from injection in the electromagnetic field illustrated in Figures 2 and 3. The local value of the number density is color coded. The spatial mesh on which the VDF is reconstructed is also shown. The figure illustrates a later stage of the interaction between the cloud and the central region of the discontinuity where the magnetic field vanishes. A significant number of protons moved in the region of positive $B_z$.

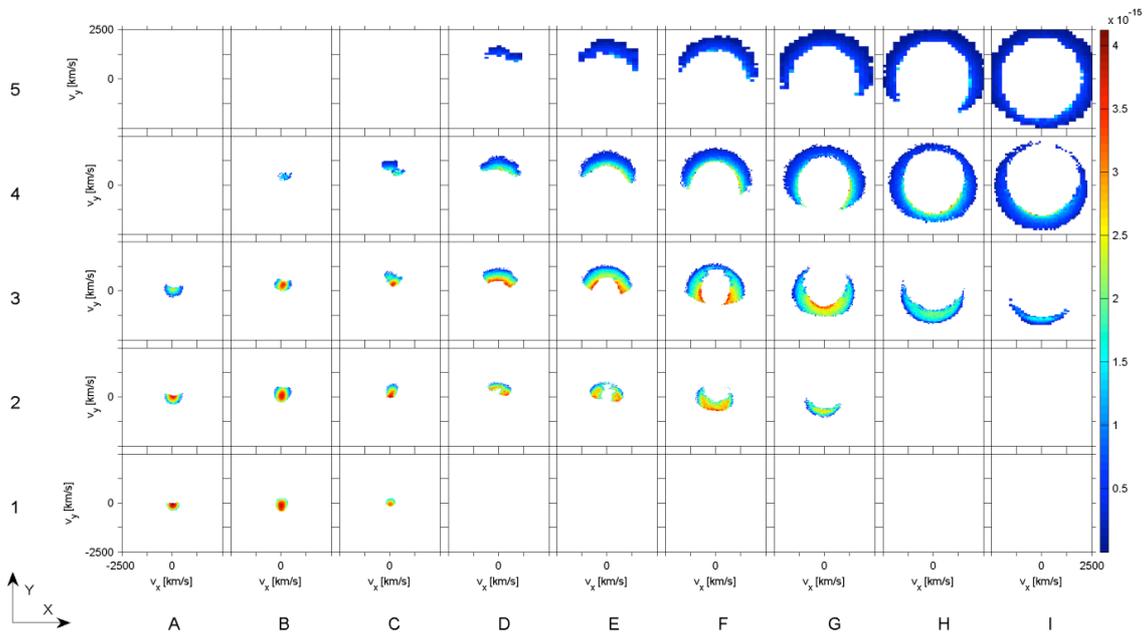

Fig. 12. Projection in the space of perpendicular velocities, for $v_z=0$, of the Liouville mapped velocity distribution functions of protons at $\Delta t=300$ seconds. The spatial bins are defined in Fig. 11. We note that in the region of positive $B_z$ the VDFs of protons are ring-shaped (bins G4–I4, H5–I5) or crescent-like (bins G3–I3, E5–G5).

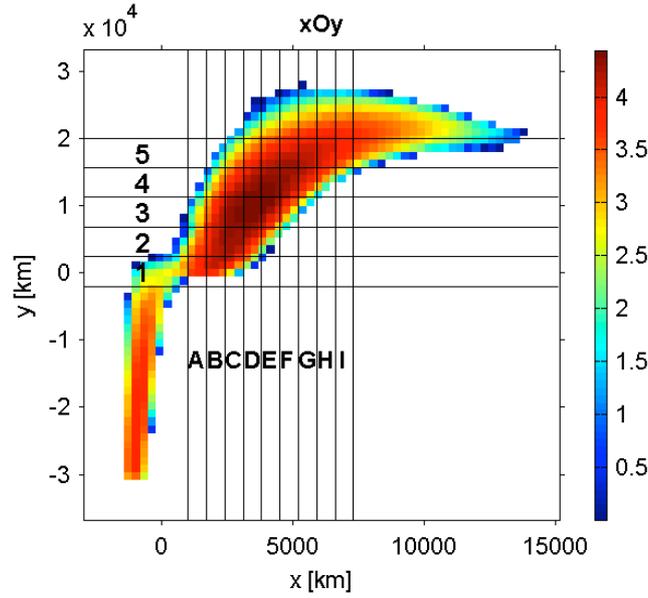

Fig. 13. Distribution of protons in the *xOy* plane after 350 seconds (~160$T_L$) from injection in the electromagnetic field illustrated in Figures 2 and 3. The local value of the number density is color coded. The spatial mesh on which the VDF is reconstructed is also shown. One notes the splitting of the cloud into two populations: population P1 that does not cross the surface where ***B***=0 and remains trapped in some region on the left side of the discontinuity (*x*<0) and respectively population P2 that penetrates into the right side of the magnetic discontinuity. At later stages the two populations disconnect. In the reminder of the paper we follow only P2.

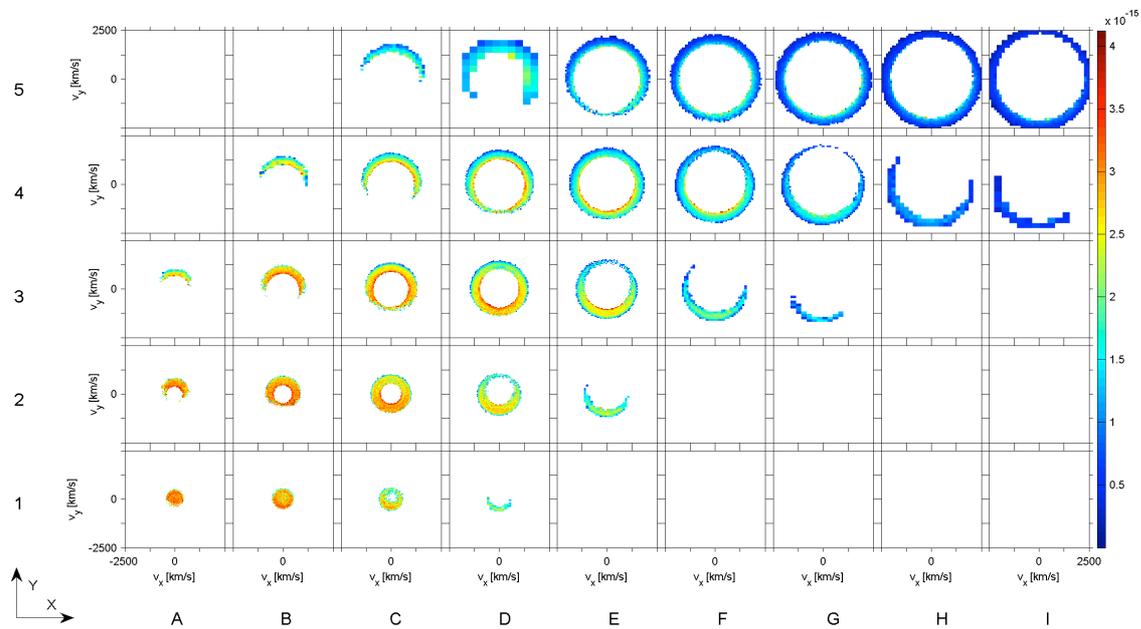

Fig. 14. Projection in the space of perpendicular velocities, for $v_z$=0, of the Liouville mapped velocity distribution functions of protons in the cloud at Δ*t*=350 seconds. The spatial bins are defined in Fig. 13. Only VDFs of the P2 population are shown. Ring-shaped and crescent-like VDFs are observed in the large majority of spatial bins.

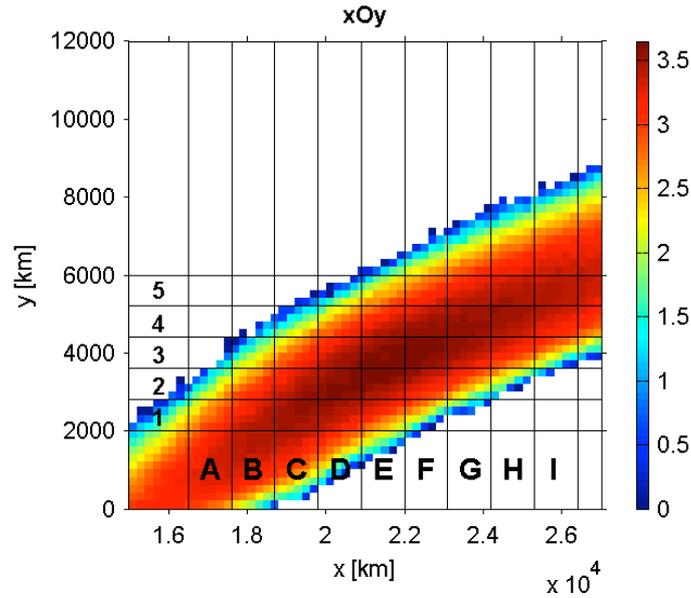

Fig. 15. Distribution of protons of the population P2 in the *xOy* plane after 600 seconds (~270$T_L$) from injection in the electromagnetic field illustrated in Figures 2 and 3. The local value of the number density is color coded. The spatial mesh on which the VDF is reconstructed is also shown. The protons move in a region of uniform magnetic and electric field, on the right side of the discontinuity.

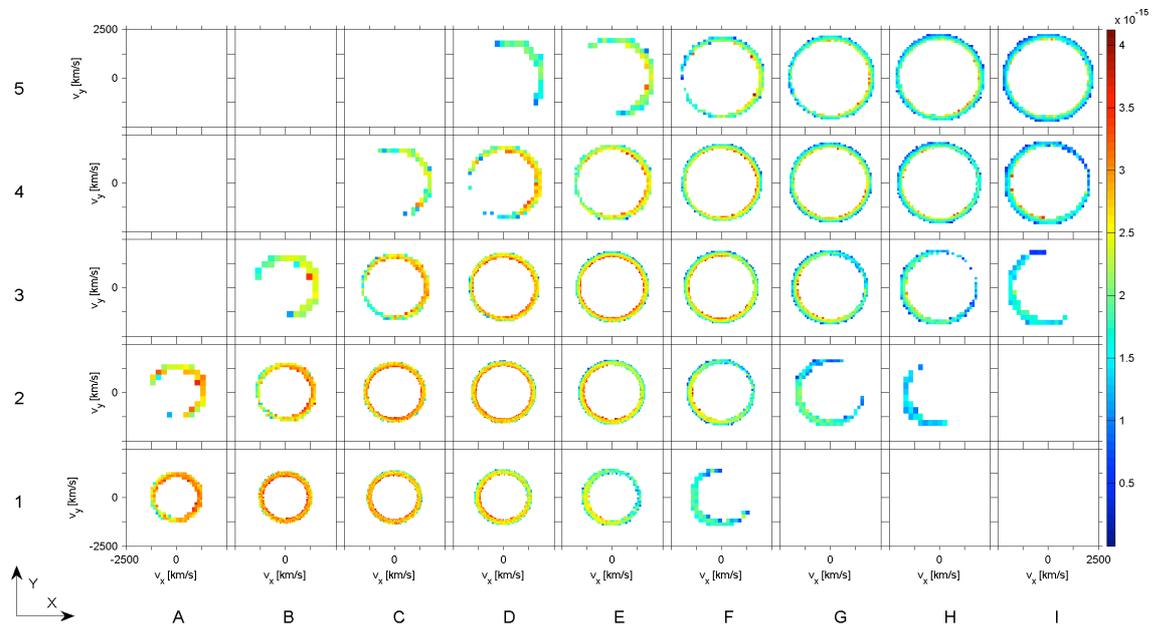

Fig. 16. Projection in the space of perpendicular velocities, for $v_z$=0, of the Liouville mapped velocity distribution functions of protons in the cloud at $\Delta t$=600 seconds. The spatial bins are defined in Fig. 15. All the VDFs obtained for this stage of propagation are either ring-shaped or crescent-like, a signature of the interaction of the cloud with the region of magnetic field gradient.

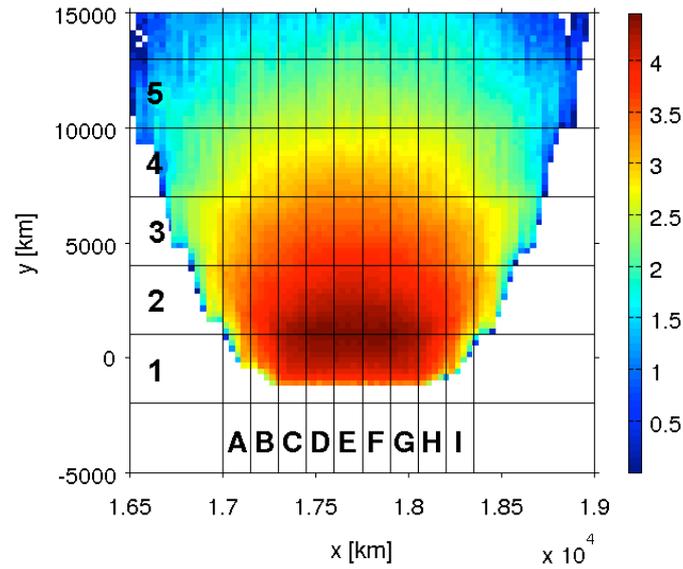

Fig. 17. Distribution of protons in the *xOy* plane after 360 seconds (~160$T_L$) from injection on the left side in the case of a unidirectional, non-uniform, parallel magnetic field and a uniform electric field. The local value of the number density is color coded. The spatial mesh on which the VDF is reconstructed is also shown. The deformation of the shape of the cloud is due to the gradient-B drift.

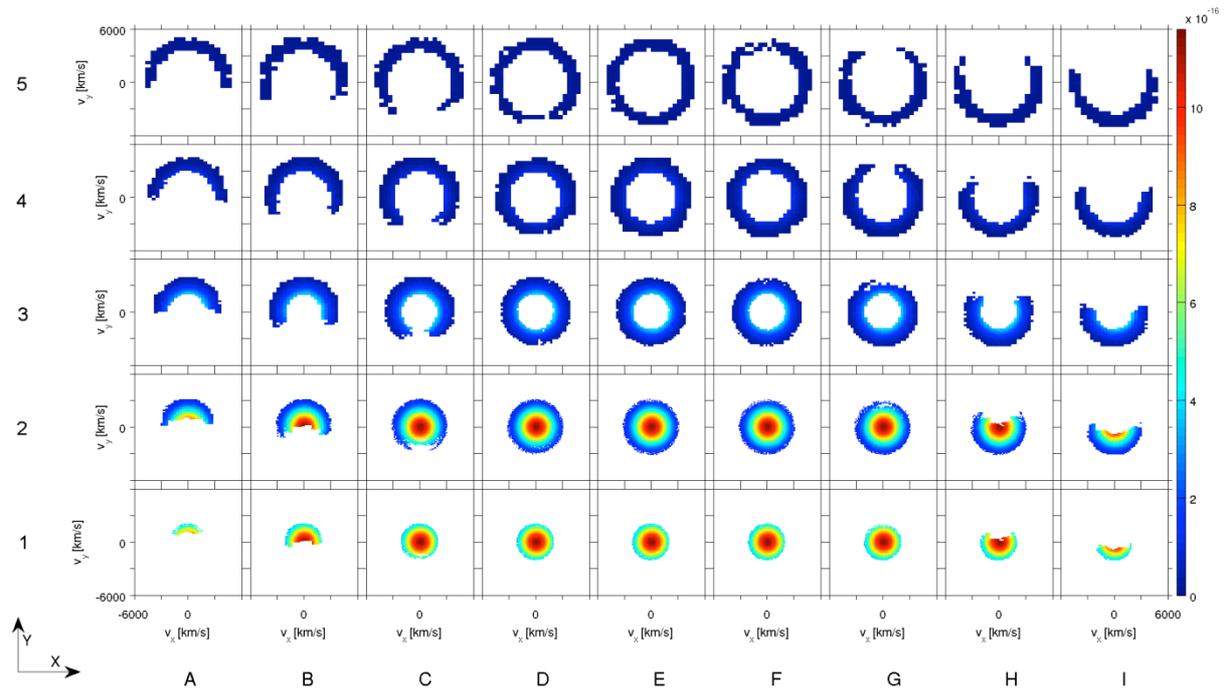

Fig. 18. Projection in the space of perpendicular velocities, for $v_z$=0, of the Liouville mapped velocity distribution functions of protons in the cloud at $\Delta t$=600 seconds for a parallel magnetic field and a uniform electric field. Spatial bins are defined in Fig. 17. Note the formation of the central cavity due to the gradient-B drift in bins of the upper three rows; non-gyrotropic VDFs are obtained in bins from the column A, B, C, G, H and I.